\documentstyle[12pt,aaspp4]{article}
\newcommand{\T}{\mathop{\rm T}\nolimits}

\begin{document}

\hfill Submitted to The Astronomy Letters.

\title{The Activity of the Soft Gamma Repeater  SGR~1900+14 \\
       in 1998 from Konus-Wind Observations:\\
       1. Short Recurrent Bursts.}

\author{E.P.Mazets\altaffilmark{1},
        T.L.Cline\altaffilmark{2},
        R.L.Aptekar\altaffilmark{1},
        P.Butterworth\altaffilmark{2},
        D.D.Frederiks\altaffilmark{1},
        S.V.Golenetskii\altaffilmark{1},
        V.N.Il'inskii\altaffilmark{1},
        V.D.Pal'shin\altaffilmark{1}
        }

\altaffiltext{1}{Ioffe Physical-Technical Institute, St.Petersburg, 194021, Russia. Mazets@pop.ioffe.rssi.ru}

\altaffiltext{2}{Goddard Space Flight Center, Greenbelt, MD 20771, USA.}

\vspace{2cm}

\begin{abstract}

Results are presented of the observations of the soft gamma repeater 
SGR 1900+14 made on the Wind spacecraft during the source reactivation 
period from May 1998 to January 1999. Individual characteristics of 
recurrent bursts, such as their time histories, energy spectra, and 
maximum and integrated energy fluxes, are considered. Some statistical 
distributions and relationships are also presented. The close 
similarity of these events to the recurrent bursts observed from other 
SGRs argues for a common emission mechanism.

\end{abstract}

\clearpage

\section{INTRODUCTION}

Recurrent short gamma-ray bursts with soft spectra have been known for 
over 20 years. The first two sources of such bursts were discovered and localized in 
March 1979 by the Konus experiment on Venera~11 and 12 (Mazets et al., 1981).
The extraordinary superintense gamma-ray outburst on March 5, 1979 
(Mazets et al., 1979a) was followed by a series of 16 weaker short bursts 
from the 
FXP~0526-66 source, which were observed during the next few years (Golenetskii et al., 1984). Also in March 1979, three short soft 
bursts arriving from the source B1900+14 were detected (Mazets et al., 1979b). In 
1983, Prognoz-9 and ICE observed a series of soft recurrent bursts from 
a third source, 1806-20 (Atteia et al., 1987, Laros et al., 1987).
The sources of recurrent soft bursts 
were given the name soft gamma repeaters, SGRs. Interestingly, a 
retrospective analysis of Venera~11 and Prognoz-7 data shows that the short 
gamma-ray burst of 07.01.1979 (Mazets et al., 1981) also belonged to SGR 1806-20 
(Atteia et al., 1987). Thus bursts from the first three soft gamma repeaters 
were detected within 
a three month period. This is remarkable, because the fourth soft gamma 
repeater, the SGR~1627-41, was detected and localized only 19 years later, 
in 1998 (Hurley et al., 1999a, Woods et al., 1999). 
A fifth SGR has also been observed (Hurley et al., 1997), but it still 
awaits a good localization.

Important new results came from studies aimed at association of the 
recurrent bursters with astrophysical objects visible in other 
wavelengths. The giant narrow initial pulse of the 1979 March 5 event was 
detected on a dozen different spacecraft. Triangulation yielded a 
very small source-localization box, about 0.1 square-arcmin, which 
projected on the outer edge of the N49 supernova remnant in the Large 
Magellanic Cloud (Cline et al., 1982). Later, ROSAT found a persistent 
X-ray source in this region (Rothschild et al., 1994).

The association of the SGR 0526-66 with N49 was sometimes questioned because of
energy considerations (Mazets and Golentskii, 1981).
For a distance of 55~kpc to 
the N49, the energy release in the March~5 event is $5 \times 10^{44}$~erg,
and in the 
recurrent bursts, up to $8 \times 10^{42}$~erg, giving luminosities which 
exceed the 
Eddington limit for a neutron star by a factor of $10^4 - 10^6$ 
(Mazets and Golenetskii, 1981). 
However arguments for large distances to the SGRs and, accordingly, for 
large energy releases continued to accumulate. Kulkarni and Frail (1993) 
established an association of the SGR~1806-20 with the supernova remnant 
G10.0-0.3, about 14~kpc distant (Corbel et al., 1997). Murakami et al. (1994) 
used ASCA to localize one of the events from SGR~1806-20 and discovered a 
soft X-ray source coinciding with its position. Subsequently, the 
observations of Kouveliotou et al. (1998) from RXTE revealed 
regular pulsations in the emission of this source with a period $P=7.47$~s.
A parallel analysis of ASCA archived data for 1992 confirmed this 
period and permitted determination of its derivative $\dot{P}=2.6\times 10^{-3}$~s~yr$^{-1}$.

SGR~1900+14 is located close to the G48.2+0.6 supernova remnant and 
is believed to be associated with it (Kouveliotou et al., 1994).
Coinciding with this repeater in position is a 
soft X-ray source reliably localized from ROSAT (Hurley et al., 1996).
ASCA observations of this source made in April 1998 revealed a 5.16-s 
periodicity of the emission (Hurley et al., 1999b).
When the SGR~1900+14 resumed its 
activity in June and August 1998 (Hurley et al., 1999c), RXTE observations 
confirmed this period and established the spin-down rate of the neutron 
star $\dot{P}=3.5\times 10^{-3}$~s~yr$^{-1}$ (Kouveliotou et al., 1999).

RXTE observations also yielded some evidence for a possible 6.7-s 
periodicity of the new SGR~1627-41 (Dieters et al., 1998). Thus the known 
soft gamma repeaters exhibit an association with young ($<10^4$~years) supernova 
remnants, a periodicity of 5-8~s, and a secular 
spin-down by a few ms per year. Thompson and Duncan (1995, 1996) suggested 
that the soft gamma repeaters are young neutron stars with superstrong, up 
to $10^{15}$~G, magnetic fields and high spin-down rates because of high losses 
due to magnetic dipole radiation -- the so-called magnetars. The fractures 
produced by magnetic stress in a neutron star's crust give rise to the release and 
transformation of magnetic energy into the energy carried away by 
particles and hard photons.

In this paper, we present the results of 
observations of recurrent bursts from SGR~1900+14 made in 1998 with a gamma-ray 
burst spectrometer onboard the Wind spacecraft (Aptekar et al., 1995).

\section{OBSERVATIONS}

Until 1998, recurrent bursts from the SGR~1900+14 were observed 
during two intervals:
three events in 1979 (Mazets et al., 1979b) and another three 
in 1992 (Kouveliotou et al., 1993). SGR~1900+14 resumed burst 
emission in May 1998 (Hurley et al., 1998; Hurley et al., 1999d), which 
continued up to January 1999.

This time the frequency of recurrent bursts was found to be high and very 
irregular. Figure~1 shows the distribution within this time interval of 
the recurrent bursts with measured fluences $S$. Three 
subintervals with a distinctly higher source activity stand out. 
On August~27, 1998, SGR~1900+14 emitted a superintense outburst with a complex and 
spectacular time structure (Cline et al., 1998; Hurley et al., 1999d).
This event is not shown in Fig.~1 because it will be considered in 
a separate paper (Mazets et al., 1999a). On May~30, 1998, an intense train of 
recurrent bursts occurred. Several tens of bursts varying in duration from 
0.05 to 0.7~s arrived during as short a time as three minutes. The 
intervals between the bursts decreased at times to such an extent as to 
become comparable to the duration of the bursts themselves, and the 
radiation intensity between them did not drop down to the background 
level. Figure~2 displays the most crowded part of the train. In Fig.~1 it 
is represented by the feature with total flux $S = 5.8 \times 10^{-5}$~erg~cm$^{-2}$.
The high burst 
occurrence frequency may cause losses in the information obtained. 
Readout of the information on a trigger event takes up about one hour. If 
other events arrive within this interval, only a very limited amount of 
the relevant information will be recorded in the housekeeping channel. 
Such cases did occur, and quite possibly they comprised two or three 
weaker trains of recurrent bursts, in particular, on September~1, 1998, 
61232.--61585~s UT, with a total flux $S \sim 2 \times 10^{-5}$~erg~cm$^{-2}$,
and on October~24, 
1998,  4921--5348~s UT, with $S \sim 10^{-5}$~erg~cm$^{-2}$. 

All recurrent bursts are short events with a fairly complex time structure 
and soft energy spectra, which, when fitted with a $dN/dE\propto E^{-
1}\exp(-E/kT)$ relation, are characterized by $kT \simeq 20-30$~keV. 
Figure~3 presents time histories of several events recorded before August~27.
Their energy spectra are very similar. Figure~4 shows the 
spectrum of a burst on June~7. After the August~27 event, the second 
interval of increased activity began (see Fig.~1),
but most of the bursts did not change their characteristics.
Shown in Fig.~5 are time structures 
of a few events, and Fig.~6 displays a typical energy spectrum. 
The only pronounced difference in the period after August~27 was the onset 
of several long, up to 4~s, bursts with a correspondingly high total 
energy flux, up to $5 \times 10^{-5}$~erg~cm$^{-2}$. Figure~7 presents time histories of two 
such bursts, with a typical energy spectrum shown in Fig.~8. Such long 
recurrent bursts were observed to be produced by other SGRs as well 
(Golenetskii et al., 1984).

As can be seen from these data, recurrent bursts exhibit a complex time 
structure, which cannot be described by a model of a single pulse with 
standard characteristic rise and decay times. The burst intensity rises 
in 15-20~ms. By contrast, long bursts take a substantially longer time to 
rise, up to $\sim 150$~ms (Fig.~7). In many cases the main rise is preceded by 
an interval with a weaker growth in intensity or even by a single weak 
pulse (Figs.~3 and 7). The intensity decay extends practically through the 
whole event. At the end of a burst one frequently observes a strong 
steepening of the falloff (Figs.~3 and 7). Large-scale details in the time 
structure may indicate that the bursts consist of several 
structurally simpler but closely related events (Figs.~3, 5, and 7).

The value of $kT$ for the photon spectra of different bursts lies in the 
18-30~keV region. There is practically no spectral evolution 
within any one event, which is readily seen from Fig.~7. The maximum 
fluxes in a burst vary from $2 \times 10^{-6}$ to $5 \times 10^{-5}$~erg~cm$^{-2}$~s$^{-1}$.
However for 80\% of events 
they lie within a narrow region of $(1-3) \times 10^{-5}$~erg~cm$^{-2}$~s$^{-1}$.
Fluences vary within broader ranges, from $10^{-7}$ to $5 \times 10^{-5}$~erg~cm$^{-2}$.
This implies that 
the energy release is partially determined by the duration of the emission process in 
the source. Figure~9 presents a fluence vs duration distribution 
of bursts $(\lg S \; vs \; \lg \Delta \T_{0.25})$.
The measure of burst duration $\Delta \T_{0.25}$ is the 
time interval within which the radiation intensity is in excess of the 25\% 
level of the maximum flux Fmax. The graph demonstrates a strong 
correlation between these quantities ($\rho = 0.8$).

As follows from the data presented here, for a 10~kpc distance of the 
SGR~1900+14 source (Case and Bhattacharya, 1998; Vasisht et al., 1994),
 and assuming the emission to be 
isotropic, the maximum source luminosity in recurrent bursts lies within 
the range of $(1-4) \times 10^{41}$~erg~s$^{-1}$, and the energy liberated in a recurrent 
burst is $2 \times 10^{39}$ to $8 \times 10^{41}$~erg.

\section{CONCLUSION}

The observations of this period of high activity of SGR~1900+14 have 
substantially broadened our ideas concerning such sources. 
The giant outburst on August~27 has come as a real surprise. Among other remarkable 
events is the intense train of bursts on May~30, 1998, when the 
frequency of recurrent bursts increased within a few minutes by at least 
a factor of $10^4$ compared to that usually observed during the reactivation periods of 
known SGRs. On the other hand, the characteristics of the bursts 
themselves, their time histories, spectra, and intensity do not suggest 
radical differences from those of recurrent events observed in other soft 
gamma repeaters (Kouveliotou et al., 1987; Frederiks et al., 1997; Mazets et al., 1999b),
which argues for the fundamental similarity between the emission processes occurring in 
different sources. It appears significant that the giant outburst with an 
energy release thousands of times larger than that typical of a single 
recurrent event did not noticeably affect the behavior and 
individual characteristics of recurrent bursts.

\acknowledgments
Partial support of the RSA and RFBR (Grant 99-02-17031) is gratefully 
acknowledged.

\clearpage

\section*{FIGURE CAPTIONS}
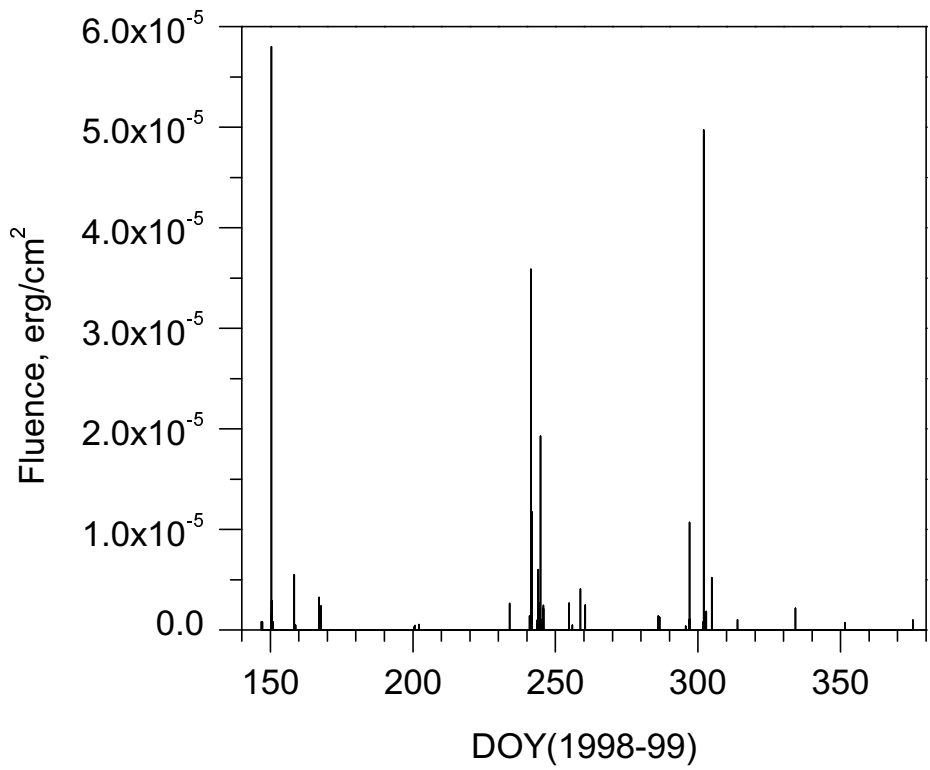
\figcaption{Distribution of the recurrent-burst occurrence and fluences during 
the source reactivation period in 1998.} 
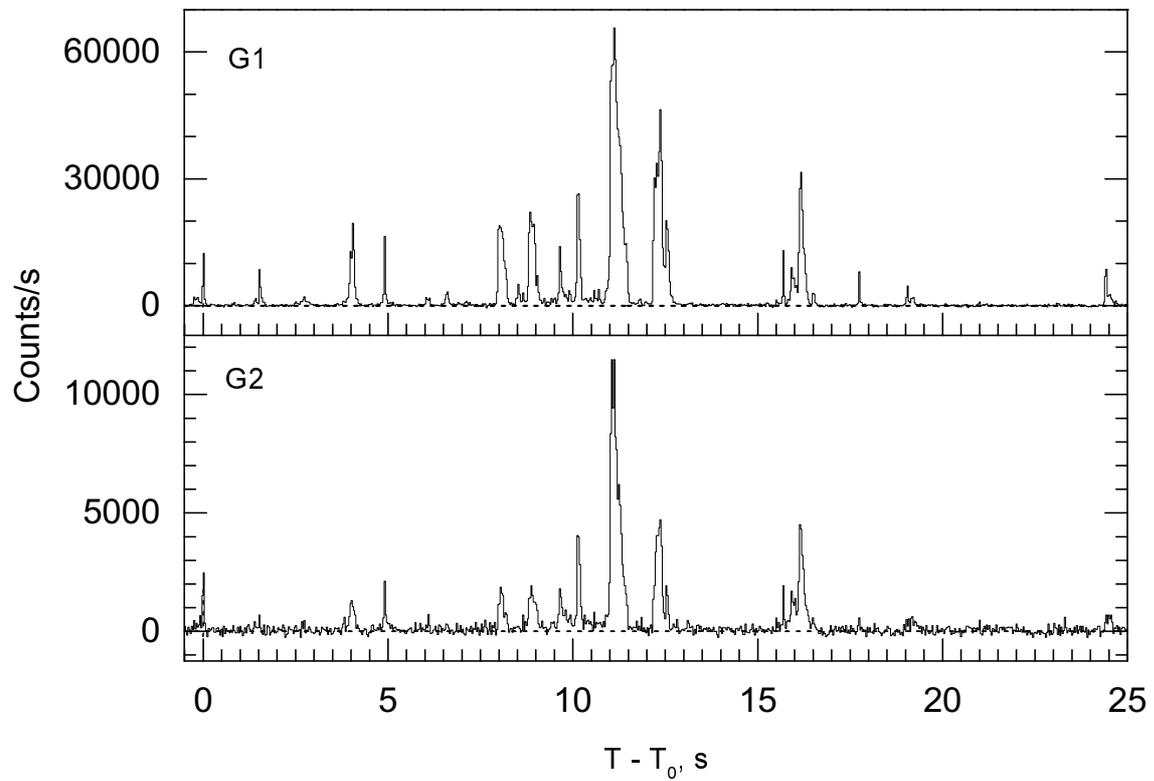
\figcaption{The interval with the highest occurrence frequency in the 
recurrent-burst train on May 30, 1998. G1 and G2 display 15--50~keV and 50--250~keV 
backgroud-substructed count rates, respectively.} 
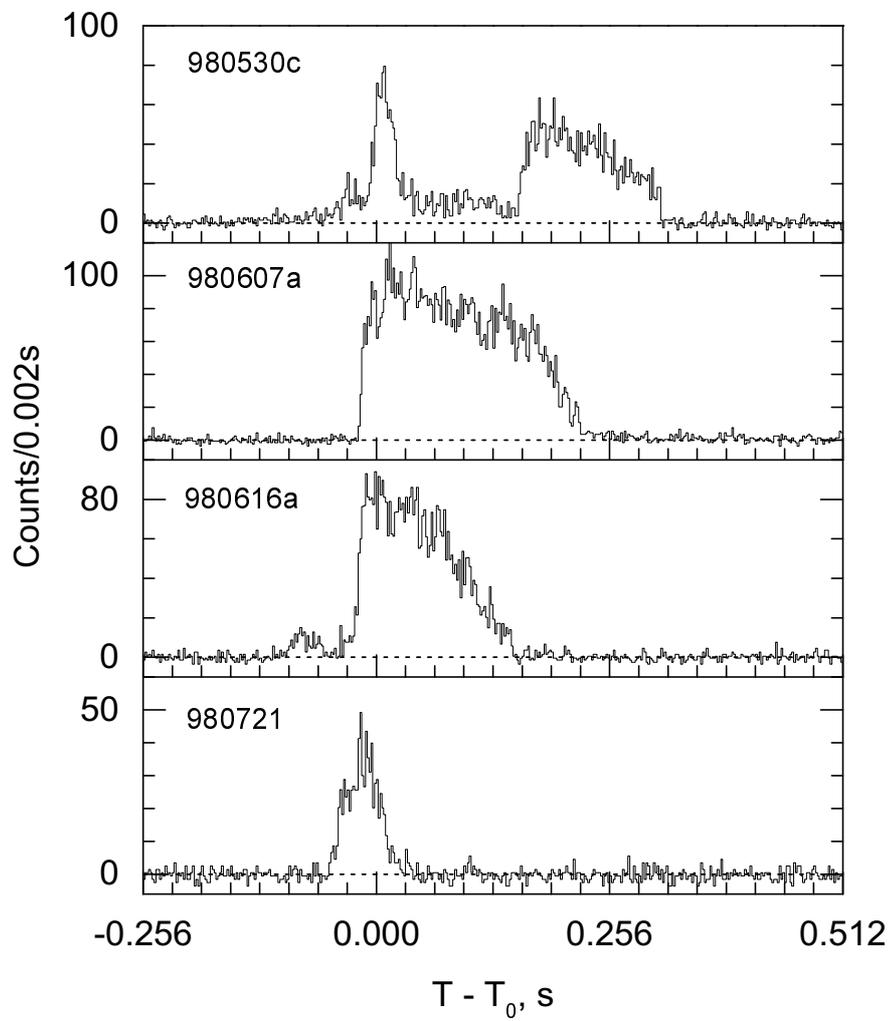
\figcaption{Time structure of several bursts recorded before the August 27 
giant burst.} 
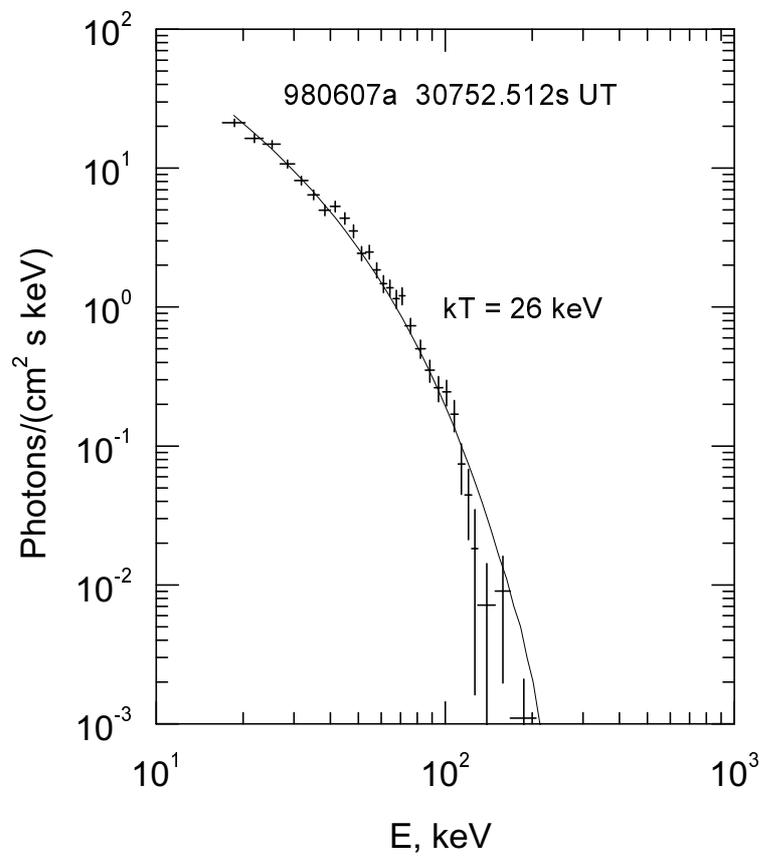
\figcaption{A typical energy spectrum of one of the bursts 
displayed in Fig.~3 (980607a).} 
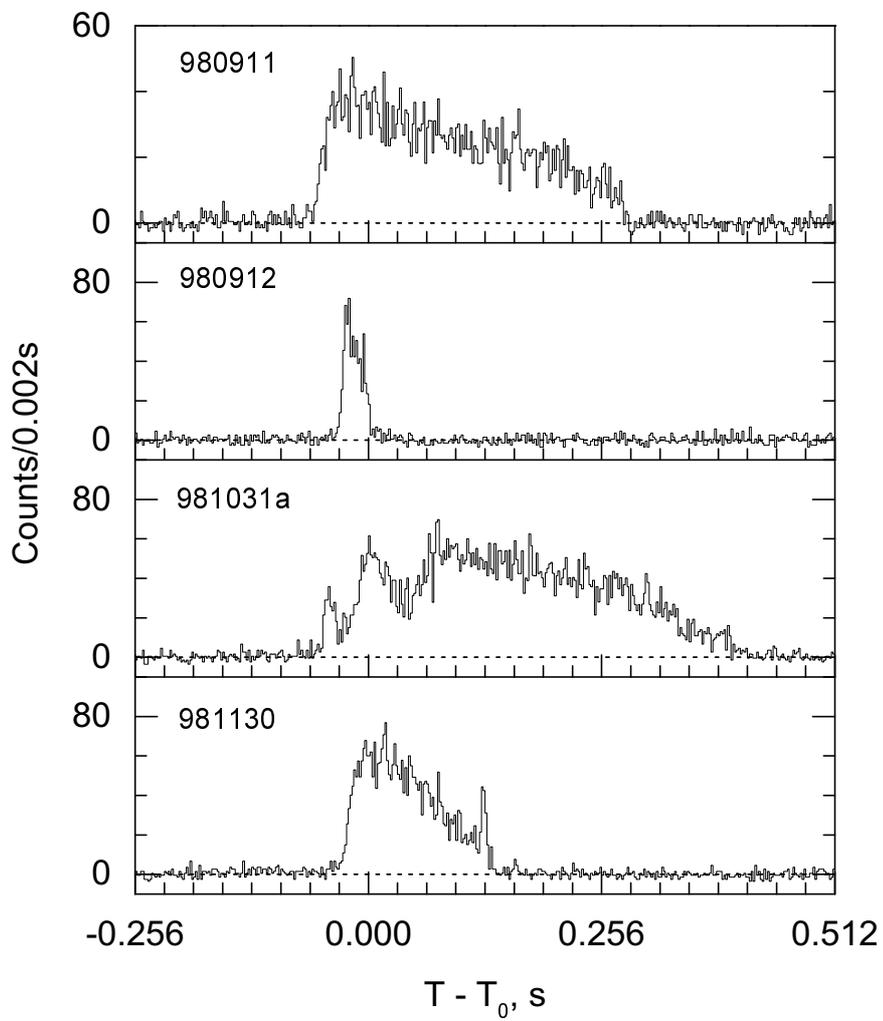
\figcaption{Time structures of bursts observed after the August 27 event.} 
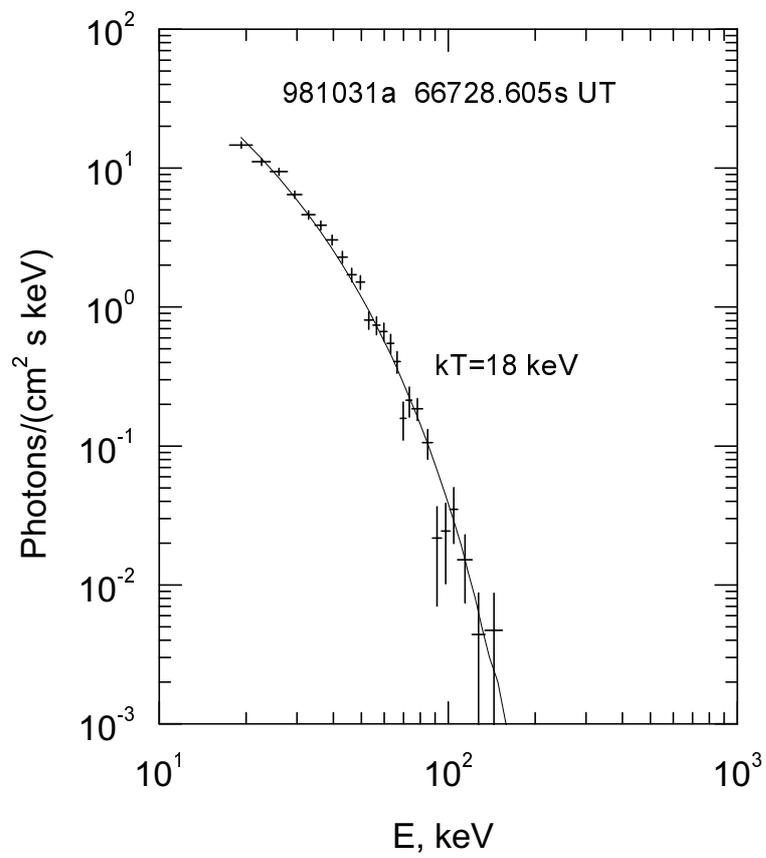
\figcaption{A typical energy spectrum of one of the bursts (Fig. 5, 981031a).} 
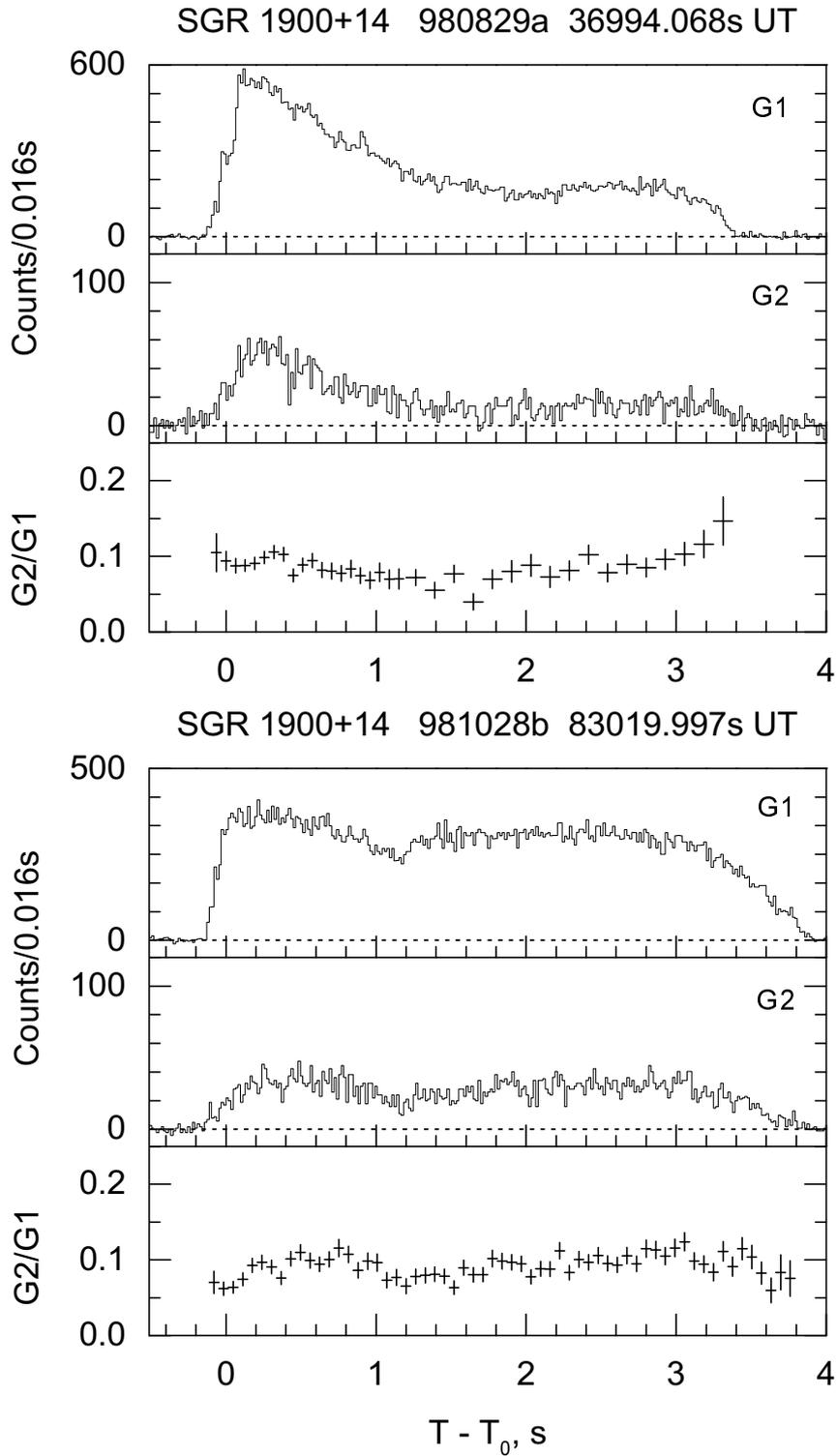
\figcaption{Time histories of two long recurrent bursts recorded in the 15--50 
and 50--250-keV energy windows. The behavior of the count-rate ratio of 
these windows, which characterizes the spectral rigidity, does not 
practically reveal any spectral evolution in the SGR 1900+14 bursts.} 
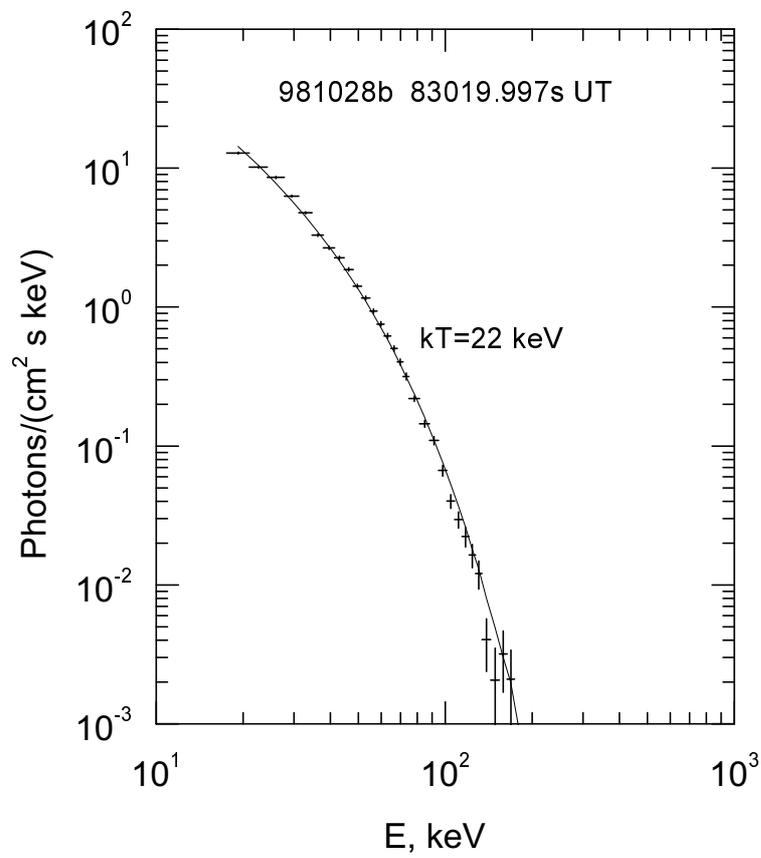
\figcaption{The spectrum of the 981028b burst.} 
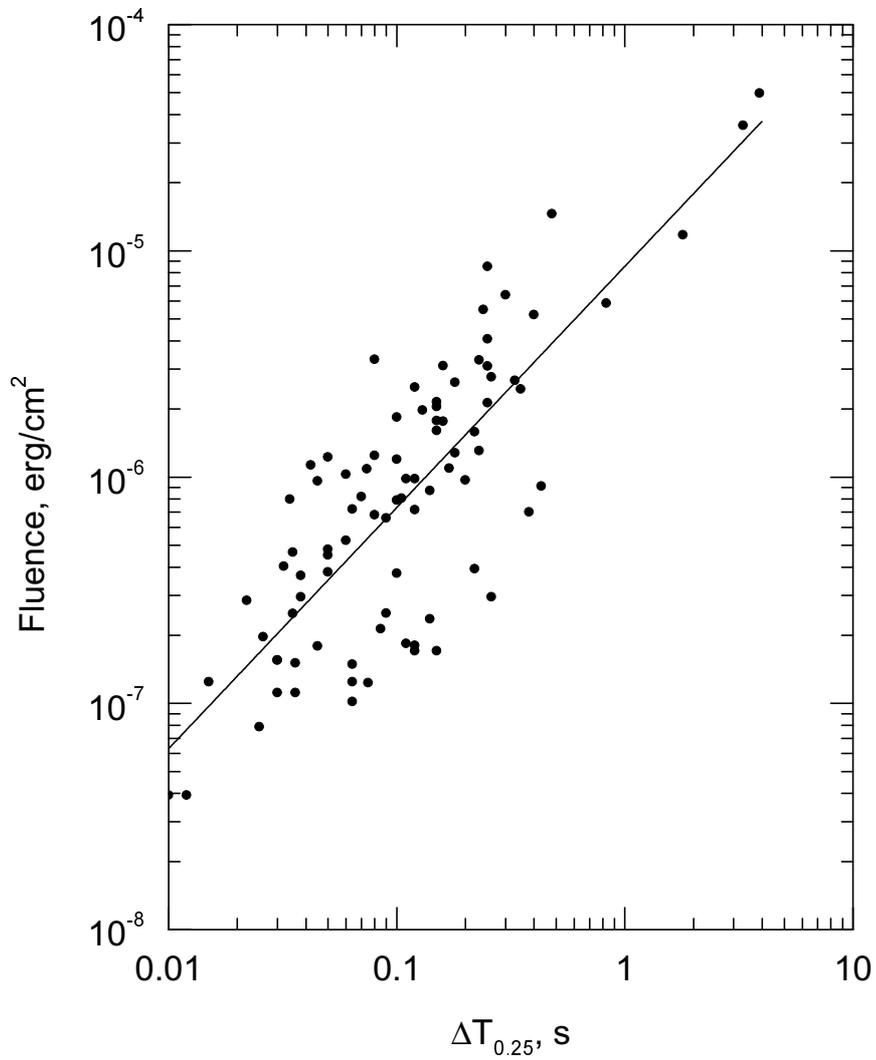
\figcaption{The strong correlation between the fluence and duration of 
a burst implies that the energy release in a source is proportional to the 
duration of emission for a small luminosity scatter.} 

\end{document}